\documentclass{article}
\usepackage{arxiv}

\usepackage[utf8]{inputenc} 
\usepackage[T1]{fontenc}    
\usepackage{hyperref}       
\usepackage{url}            
\usepackage{booktabs}       
\usepackage{amsfonts}       
\usepackage{nicefrac}       
\usepackage{microtype}      
\usepackage{lipsum}		
\usepackage{graphicx}
\title{Accelerating Geometric Multigrid
Preconditioning\\ with Half-Precision Arithmetic on GPUs}

\author{ $\text{Kyaw L. Oo}$,$\,$ $\text{Andreas Vogel}$ \\
High Performance Computing in the Engineering Sciences\\
Department of Civil and Environmental Engineering\\
Ruhr University Bochum, Germany \\
 \{kyaw.oo, a.vogel\}@rub.de
}

\hypersetup{
pdftitle={A template for the arxiv style},
pdfsubject={q-bio.NC, q-bio.QM},
pdfauthor={David S.~Hippocampus, Elias D.~Striatum},
pdfkeywords={First keyword, Second keyword, More},
}
\usepackage{cite}
\usepackage{amsmath,amssymb,amsfonts}
\usepackage{algorithmic}
\usepackage{graphicx}
\usepackage{textcomp}
\usepackage{xcolor}
\usepackage{pgfplots}
\definecolor{aureolin}{rgb}{1, 0.55, 0.0}
\usetikzlibrary{pgfplots.groupplots}
\usepackage{hyperref}
\hypersetup{
  colorlinks   = true, 
  urlcolor     = blue, 
  linkcolor    = blue, 
  citecolor    = blue  
}
\usepackage[norelsize, linesnumbered, ruled, noline, boxed, commentsnumbered]{algorithm2e}
\usepackage{booktabs} 

\usepackage{tikz}
\usepackage{pgfplots}
\usepackage{siunitx}
\usepackage[para,online,flushleft]{threeparttable}

\newcommand{\func}[1]{\texttt{#1}}

\begin{document}
\maketitle

\begin{abstract}
With the hardware support for half-precision arithmetic on NVIDIA V100 GPUs, high-performance computing applications can benefit from lower precision at appropriate spots to speed up the overall execution time. 
In this paper, we investigate a mixed-precision geometric multigrid method to solve large sparse systems of equations stemming from discretization of elliptic PDEs. 
While the final solution is always computed with high-precision accuracy, an iterative refinement approach with multigrid preconditioning in lower precision and  residuum scaling is employed. 
We compare the \func{FP64} baseline for Poisson's equation to purely \func{FP16} multigrid preconditioning and to the employment of \func{FP16}-\func{FP32}-\func{FP64} combinations within a mesh hierarchy. 
While the iteration count is almost not affected by using lower accuracy, the solver runtime is considerably decreased due to the reduced memory transfer and a speedup of up to $2.5\times$ is gained for the overall solver. 
We investigate the performance of selected kernels with the hierarchical Roofline model.
\end{abstract}

\keywords{Half-precision, Elliptic Partial Differential Equation, Geometric Multigrid Method, Mixed-precision Iterative Refinement, GPU, Sparse Linear Algebra}

\section{Introduction}
\par 
\par 
Floating-point calculations are omnipresent in scientific computing and  consequently broadly supported on computing architectures. If several precision formats are supported on a device, this can be exploited to accelerate numerical algorithms by lowering the accuracy whenever admissible within an algorithm. A speedup can then be gained due to reduced data transfer in memory-bound applications or due to a higher peak performance at lower precision in the compute-bound cases. For example, the NVIDIA V100 GPU achieves up to 112 TFLOP/s in half precision while only 7 TFLOP/s in double precision and 14 TFLOP/s for single-precision floating-point operations. However, an algorithmic reformulation is often necessary to allow for a beneficial employment of lower precision without sacrificing the final result accuracy.

In scientific computing, elliptic partial differential equations (PDE) are frequently encountered to model the behavior of physical systems.
For instance, Poisson's equation has numerous applications in fluid dynamics and electromagnetism. 
Approximation techniques such as the finite difference method (FDM) or the finite element method (FEM)~\cite{Brenner94} are then commonly used to obtain a numerical solution of the PDE.
These grid-based discretization techniques lead to large sparse linear systems of equations in the form of $\mathbf{Au=b}$, where $\mathbf{A}$ is the sparse stiffness matrix, $\mathbf{u}$ the discrete solution vector and $\mathbf{b}$ a source term vector. A significant amount of time is typically spent to find a solution for such matrix systems, i.e., to find an approximate inverse of the stiffness matrix.
The multigrid method~\cite{Hackbusch85} has been shown to be one of the most efficient methods for these kind of problems. It is composed of simpler algorithmic kernels combined in a suitable way over several discretization levels and features a linear complexity as well as a bounded iteration count independent of the grid resolution.

In this contribution, we investigate the benefit of the employment of half-precision accuracy on the V100 GPU for the geometric multigrid method. As representative application, Poisson's equation is solved employing an iterative refinement~\cite{Wilkinson1963} algorithm with a low-precision multigrid preconditioning and the gained speedup over a purely double-precision implementation is measured. We employ the IEEE \num{754} standard \func{FP16} accuracy which occupies \num{16}-bits  (\num{1} sign bit, \num{5} exponent bits, \num{10} fractional bits) and has a dynamic range resulting in relatively high precision for floating-point values near zero, but low precision for values far away from zero. This property makes it suitable to be used in scientific computation, however, the limited range of \func{FP16} (see Table~\ref{tab:my-table}) requires to use a
scaling factor, similar to the ones used in deep learning training, to keep the floating-point values in the representable range of \func{FP16} throughout the algorithm.

\begin{table}[!h]
\centering
\begin{threeparttable}
\caption{IEEE standard 754 floating-point precision}
\label{tab:my-table}
\begin{tabular}{@{}llll@{}}
\toprule
 & $r_\text{min}\tnote{1}$ & $r_\text{max}$\tnote{2} & Unit Roundoff \\ \midrule
\func{FP16}  & $ 6.10 \times 10^{-5}$ & $ 65504 $  & $2^{-11} \approx 4.9 \times 10^{-4} $ \\ 
\func{FP32}  & $ 1.18 \times 10^{-38}$ & $ 3.4\times 10^{38}$&  $2^{-24} \approx 6.0 \times 10^{-8} $ \\ 
\func{FP64} & $ 2.22 \times 10^{-308}$ & $ 1.80 \times 10^{308}$ & $2^{-53} \approx 1.1 \times 10^{-16} $  \\  
\end{tabular}
\begin{tablenotes}\footnotesize
\item [1] minimum  positive (normal) value \item [2] maximum representable value
\end{tablenotes}
\end{threeparttable}
\end{table}


\section{Related Work}
Iterative refinement (\func{IR}) \cite{Wilkinson1963,Martin1966203,Moler1967316} is a popular technique to iteratively solve a linear system of equations and can be broken down into three consecutive  tasks for each iteration step: 
residual computation for a given iterate, solving a correction equation for the residual, and adding the correction to the current iterate. 

\func{IR} dates back to Wilkinson in the 1940s and a lot of improvements and mathematical investigations have been developed since then. Comprehensive overviews and references can be found, e.g., in \cite{Stewart1973,Demmel2006325,Higham2002}. 
Remarkably, all three steps of \func{IR} can be performed in a different precision and the correction equation only has to be solved approximately.
A recent mathematical analysis for the convergence properties of a three precision scenario can be found in \cite{Carson2017A2834,Carson2018A817}.
The employment of iterative correction equation solvers, also called inner solvers, has been studied for PCG~\cite{Golub19991305},  GMRES~\cite{Turner1992815,Saad1991}, and Krylov subspace methods~\cite{Simoncini20022219}.

\par
In 2006, motivated by the speedup gains of using single-precision over double-precision floating-point accuracy, dense linear algebra has been accelerated employing LU factorization in \func{FP32} within an \func{FP64} iterative refinement~\cite{Langou2006,denselinear}. The employment of an inner \func{FP32} GMRES solver has been studied in~\cite{Baboulin20092526}.



For FEM simulations and employing geometric multigrid, iterative refinement and \func{FP32} precision preconditioning for an outer \func{FP64} solution accuracy has been extensively studied in~\cite{Goeddeke2007221}. A residuum scaling heuristic has been employed for the iterative refinement and different convergence criteria for the inner multigrid solver, such as fixed iteration count or residual norm reduction, have been investigated. The authors observe that outer \func{FP64} accuracy is mandatory for final FEM solutions, but considerable speedup is gained by employing \func{FP32} multigrid on the GPU as inner solvers for the \func{IR}. A cascading mixed-precision multilevel solver has been discussed but not studied due to the lack of available hardware at the time. 
In~\cite{goddek}, it was summarized that mixed-precision iterative refinement multigrid schemes are always more efficient than using double precision exclusively, and the application of cyclic reduction smoothers for mixed-precision multigrid has been presented in~\cite{GoeddekeCyclicReduction}. The mixed-precision idea for geometric multigrid using single-precision preconditioning has been employed subsequently in a couple of works, e.g., in~\cite{osborn2010multigrid,YAMAGISHI20161658,Kronbichler2019}. Using multigrid algorithms for lattice quantum chromodynamics (LQCD)  on GPUs can be found on QUDA library \cite{qcd}. Multigrid method with single precision is used as a preconditioner with general conjugate residual outer solver in double precision.
Mixed-precision algebraic multigrid approaches using single- and double-precision floating point formulations have been investigated in~\cite{EMANS2010175,Sumiyoshi2014,amg}.

%

With the advent of half-precision hardware support on GPUs, \func{FP16} accuracy has been used to accelerate dense linear solvers in~\cite{half}.
Computing a final solution in \func{FP64}, a speedup of about $1.7\times$ has been reported for \func{FP32} inner solver, and a speedup up to $2.7\times$ employing \func{FP16}. In~\cite{TensorCore}, half-precision arithmetic achieved a $4\times$ speedup employing Tensor Cores (\func{FP16\-TC}) on a NVIDIA V100 GPU.
Recently, a dynamic precision change between the available precisions on Volta GPUs has been presented for algebraic multigrid in~\cite{Fevre2018}.

In contrast to these works, we are going to investigate the benefit of \func{FP16} employment and cascading precisions within geometric multigrid cycles which, to the best of our knowledge, has not been studied in literature so far.


\par


\pagebreak
\section{Contribution}
\par
The main contribution of this paper is to investigate the performance gain for half-precision floating-point accuracy in geometric multigrid preconditioning on NVIDIA V100 GPUs.
In particular, we:
\begin{itemize}
    \item compare \func{FP64} preconditioning to \func{FP16} and mixed \func{FP16}-\func{FP32}-\func{FP64} geometric multigrid cycles in terms of iteration counts and runtime,
    \item show that an elliptic PDE solver can be accelerated up to $2.5\times$ with half-precision multigrid iterative refinement compared to purely double-precision multigrid while maintaining the same solution accuracy,
    \item provide a Roofline model analysis of the frequently used kernels to highlight their limitations.
    \end{itemize}


\section{Methods}
Our solver is going to be a nested solver that combines the iterative refinement (\func{IR}) as outer solver with one cycle of the lower-precision geometric multigrid to approximate the solution of the correction equation.

\subsection{Iterative Refinement Method}
The classical \func{IR} is a three-step process in which the linear system of equations $\mathbf{Au=b}$ is solved by  iteratively improving the current solution approximation $\mathbf{u}_i$ until convergence.
Given an initial guess $\mathbf{u}_0$, the steps are:
\begin{enumerate}
    \item Residuum computation: $\mathbf{r=b-Au}_i$
    \item Solving correction equation: $\mathbf{Ac=r}$
    \item Solution update:  $\mathbf{u}_{i+1}=\mathbf{u}_i+\mathbf{c}$
\end{enumerate}
\par
If the correction equation is solved exactly, the solution can be computed in one iteration assuming infinite accuracy. 
However, it is only possible to obtain the correction approximately because of round-off errors and the solution thus has to be iteratively improved until the desired accuracy is reached. 
As correction equation solver in \func{IR}, usually LU factorization with partial pivoting or  Krylov subspace methods such as the generalized minimal residual method (GMRES) and the conjugate gradient method (CG) are employed.
For the mixed-precision approach, the correction is computed in lower precision. The relative speedup gained by using lower precision depends on the hardware, the properties of  matrix and solver configurations. With appropriate lower-precision hardware support, one iterate is expected to be faster, but the iteration count might increase to reach a prescribed solution accuracy due to the less accurately computed correction.
In our study, we employ one multigrid cycle to approximate the solution of the inner correction equation.

\subsection{Multigrid Method}


In order to compute a solution on a fine grid $\Omega_L$, the multigrid method employs a hierarchy of grid levels $\{\Omega_0,\Omega_1,...,\Omega_l,...,\Omega_L\}$ with decreasing grid resolution and increasing number of unknowns. An initial guess is iteratively corrected by appropriately exploiting the different mesh resolutions.   
Roughly speaking, high error frequencies are addressed on fine grids applying smoothing (or relaxation) operations. The residuum can then be transferred to coarser grids where lower frequencies can be considered higher and are again removed by smoothing. This way, different parts of the error spectrum are addressed on different levels. Technically, the multigrid method works by recursively applying the two-grid method on the mesh levels as shown in Algorithm~\ref{twogrid}.
In order to end the recursion, an exact solve is applied at the coarsest level also called base level. 
Detailed explanations of the multigrid method can be found in~\cite{Hackbusch85,simbook}. 
If the coarse grid problem (Alg.~\ref{twogrid}, line 3) is approximated by a single step of a two-level method on the coarser level, the resulting algorithmic pattern is called V-cycle. See Fig.~\ref{dsh_mg} for an illustration.
%
%
 \begin{algorithm}[tbp]
 \caption{Two-Grid Algorithm $\mathbf{A}^{\scriptstyle \func{FP}_{\scriptscriptstyle \func{L}} }_{L}\mathbf{u}^{\scriptstyle \func{FP}_{\scriptscriptstyle \func{L}} }_{L}=\mathbf{b}^{\scriptstyle \func{FP}_{\scriptscriptstyle \func{L}} }_{L}$}
 \begin{algorithmic}[1]
 \renewcommand{\algorithmicrequire}{\textbf{Input:}}
 \renewcommand{\algorithmicensure}{\textbf{Output:}}
 \REQUIRE  $\mathbf{u}^{\scriptstyle \func{FP}_{\scriptscriptstyle \func{L}} }_L$, $\mathbf{b}^{\scriptstyle \func{FP}_{\scriptscriptstyle \func{L}} }_L$, $\nu_1$ ,$\nu_2$
 \ENSURE  $\mathbf{u}^{\scriptstyle \func{FP}_{\scriptscriptstyle \func{L}} }_L$
 \STATE On fine grid, pre-smooth $\nu_1$ times : $\mathbf{u}^{\scriptstyle \func{FP}_{\scriptscriptstyle \func{L}} }_L := S_L^{\nu_1}(\mathbf{u}^{\scriptstyle \func{FP}_{\scriptscriptstyle \func{L}} }_L)$
 \STATE  Compute residual $\mathbf{r}^{\scriptstyle \func{FP}_{\scriptscriptstyle \func{L}} }_L := \mathbf{b}^{\scriptstyle \func{FP}_{\scriptscriptstyle \func{L}} }_L - \mathbf{A}^{\scriptstyle \func{FP}_{\scriptscriptstyle \func{L}} }_L \mathbf{u}^{\scriptstyle \func{FP}_{\scriptscriptstyle \func{L}} }_L$; transfer to the coarser grid, change precision to obtain $\mathbf{r}^{\scriptstyle \func{FP}_{\scriptscriptstyle \func{L-1}} }_{L-1}$
 \STATE Solve for a coarse correction: $\mathbf{A}^{\scriptstyle \func{FP}_{\scriptscriptstyle \func{L-1}} }_{L-1}\mathbf{c}^{\scriptstyle \func{FP}_{\scriptscriptstyle \func{L-1}} }_{L-1}=\mathbf{r}^{\scriptstyle \func{FP}_{\scriptscriptstyle \func{L-1}} }_{L-1}$ 
\STATE  Transfer $\mathbf{c}^{\scriptstyle \func{FP}_{\scriptscriptstyle \func{L-1}} }_{L-1}$ to the fine grid, change precision to obtain $\mathbf{c}^{\scriptstyle \func{FP}_{\scriptscriptstyle \func{L}} }_{L}$; update the solution $\mathbf{u}^{\scriptstyle \func{FP}_{\scriptscriptstyle \func{L}} }_L := \mathbf{u}^{\scriptstyle \func{FP}_{\scriptscriptstyle \func{L}} }_L + \mathbf{c}^{\scriptstyle \func{FP}_{\scriptscriptstyle \func{L}} }_{L}$
\STATE Post-smooth $\nu_2$ times : $\mathbf{u}^{\scriptstyle \func{FP}_{\scriptscriptstyle \func{L}} }_L := S_L^{\nu_2}(\mathbf{u}^{\scriptstyle \func{FP}_{\scriptscriptstyle \func{L}} }_L)$
\STATE Test for convergence and repeat step 1 if required
 \end{algorithmic} 
 \label{twogrid}
 \end{algorithm}

Most importantly, a correction equation is solved on the next coarser meshes. This suggests that the coarse grid correction equation can be solved with a different floating-point precision than the fine grid equation (see also \cite{Goeddeke2007221}) and we indicate the different precisions as superscripts in Alg.~\ref{twogrid}. In particular, the transfer of residuum and correction has to be perform in any case and the change in accuracy can thus be introduced without an additional copy operation as given in the \func{IR} method.
For the multigrid cycle itself, multiple combinations of precisions at different grid levels can be envisioned. In Table~\ref{variants}, the most obvious choices are summarized.
\begin{table}[h]
	\centering
		\caption{Floating-point precision used in different grid levels of the mixed-precision multigrid cycle
}
	\begin{tabular}{@{}rccc@{}}
		\toprule
		Variant & Outer \func{IR} & Grid Levels (L$\rightarrow $\num{0}) & Base Solver \\ \midrule
	\textbf{\func{	D\_MG}}& FP64 & FP64 & FP64\\ \midrule
	\textbf{\func{	H\_MG}}& FP64 & FP16 & FP16 \\ \midrule
	\textbf{\func{	DSH\_MG}} & FP64 & FP64$\rightarrow$FP32$\rightarrow$FP16 & FP16 \\ \midrule
	\textbf{\func{	HSD\_MG}} & FP64 & FP16$\rightarrow$FP32$\rightarrow$FP64 & FP64 \\ \bottomrule
	\end{tabular}

	\label{variants}
\end{table}


The choice of the smoother is very important for the multigrid method.
In our study, we employ three pre- and post-smoothing steps with a damped Jacobi method with damping factor \num{2/3} for the diagonally dominant stiffness matrix $\mathbf{A}$. The Jacobi smoother requires a multiplication with the inverse diagonal which can be implemented with a simple component-wise vector multiplication. For the base solver, the CG method is employed.  
We use Jacobi smoothing because it is very simple and  easy to implement. 
While smoother like ILU or Gauss-Seidel achieve better convergence than Jacobi, they are more difficult to parallelize and advanced techniques, e.g., cyclic reduction \cite{GoeddekeCyclicReduction}, have to be used.
Since the focus of this research is to accelerate the multigrid cycle by half precision, we left aside the smoother aspect and refer to~\cite{kurzak2010scientific} for an in-depth discussion of smoother aspects. 



\subsection{Mixed-precision Iterative Refinement Method}

\begin{algorithm}[tbp]
	\SetKwComment{Comment}{$\triangleright$\ }{}
	\DontPrintSemicolon
	\KwData{ $\mathbf A_\text{high}$, $\mathbf b_\text{high}$, $\mathbf{u}_\text{high}^0$, $\epsilon$
	}
	$\mathbf A_\text{low} := \mathbf A_\text{high}$,
	$\mathbf{u}_\text{high} := \mathbf u_\text{high}^0$\;
	$\mathbf {r}_\text{high} := \mathbf b_\text{high}-\mathbf A_\text{high}\mathbf u_\text{high}$\;
	\While{$\alpha := \lVert \mathbf{r}_\text{\textrm{\textup{high}}} \rVert > \epsilon$}
	{
 		$\mathbf r_\text{low} := (1/\alpha) \, \mathbf r_\text{high}$\;
		
 		Approximate $\mathbf A_\text{low} \mathbf c_\text{low} = \mathbf r_\text{low}$ by 1 multigrid cycle 		\;
 		
 		$\mathbf{u}_\text{high}:=\mathbf u_\text{high}+\alpha\, \mathbf c_\text{low}$\;
 		$\mathbf{r}_\text{high}:=\mathbf r_\text{high}-\alpha\, \mathbf A_\text{high} \mathbf {c}_\text{low}$
	}
	\caption{Mixed-Precision Iterative Refinement with Residuum Scaling and Geometric Multigrid}
	\label{algo:mixed_IR2}
\end{algorithm}
\textbf{}
 For scientific computing applications, all the steps of \func{IR} can be computed in  double precision and the method can then be viewed as a consistent linear iterative scheme.
For the mixed-precision approach, the correction equation is computed in lower precision with the goal to make the computation faster without loosing the final solution accuracy. We employ the mixed-precision \func{IR} with multigrid preconditioning as described in Algorithm \ref{algo:mixed_IR2}. 
%
The correction equation is coarsely approximated by a multigrid cycle  using \func{FP16} precision on some or all grid levels, while the remaining part of the \func{IR} is performed at the higher \func{FP64} precision. In particular, the convergence criterion for the final solution is evaluated in double precision and thus allows to solve as accurately as with pure \func{FP64} precision provided that the \func{IR} iteration converges. Employing a Jacobi smoother, all solver components can be implemented as a series of SpMV-like and AXPY-like kernels.

One disadvantage of using \func{FP16} is its limited range.
The minimum positive normal number for \func{FP16} is $2^{-14} \approx 6.10 \times 10^{-5}$.
 For the convergence of the \func{IR} scheme,
it is  thus important to keep the floating-point values within this representable range. 
 With the progression of the \func{IR} iteration, the residuum becomes smaller with every outer iteration  since we want to solve down to a final residuum norm of $10^{-9}$ in our experiments.
The correction equation can  therefore not be solved if the high-precision residuum  is flushed to zero in a low-precision format. 
Scaling the residuum values is thus necessary to  stay within the range of \func{FP16}.
 To this end, we employ a residuum scaling in line~\num{4} of Algorithm~\ref{algo:mixed_IR2} where the cast to low precision has to be performed. By applying this scaling before precision conversion, the residuum norm becomes one, entries are thereby better distributed in the representable range, and the cast does not result in zero flushings. The scaling factor is then compensated in line 6 and 7 of the algorithm. All scalings are performed in high precision such that resulting values are suitable for the employed precision. Since the employment of lower precision for the multigrid cycle only provides a rough approximation for the correction, we employ a single cycle before recomputing the residuum in high precision.



\section{Model Problem}
\par 
The model problem considered is Poisson's equation.  For $D=2$ or $3$, given a domain $\Omega \subset \mathbb{R}^D $ with boundary $ \partial \Omega$, Poisson's equation with homogeneous Dirichlet boundary condition is given as:
\begin{align}
\begin{split}
	-\Delta u &= f , \qquad \text{in} \; \Omega \\
	u &= 0, \qquad \text{on} \;\partial \Omega,
	\end{split}
\end{align}
where $\Delta$ is the Laplace operator, $f$ is a prescribed known function and $u$ is the unknown function.
The equation  is discretized on a unit square with bilinear quadrilateral Q1 finite elements in 2D, and on a unit cube with trilinear functions in the 3D case. 
For the exact 2D solution, we choose 
\begin{equation}
u(x,y) = \sin(k\pi x) \sin(k \pi y)
\label{testFunc}
\end{equation}
with an integer parameter $k$ and enforce it by a correspondingly manufactured right-hand side $f$, and a corresponding choice in the three-dimensional case. The example has been chosen as a reproducible benchmark problem.

\par %
\section{Solver}

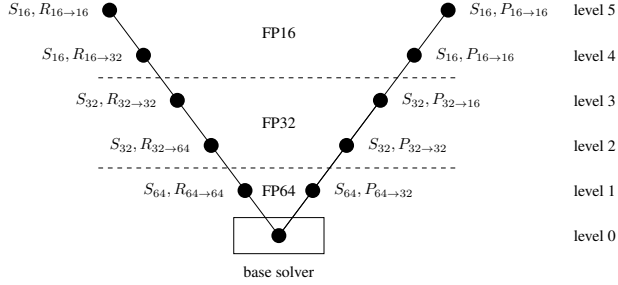
\begin{figure}[h]
 	\centering
 \scalebox{.6}{\usetikzlibrary{calc}
\begin{tikzpicture}

	\node [style=circle,fill=black] (31) at (-1., 4) {};
		\node [style=circle,fill=black] (30) at (5, 4) {};
	\node [style=circle,fill=black] (32) at (-1.75, 5) {};
		\node [style=circle,fill=black] (33) at (5.75, 5) {};
		\node [style=circle,fill=black] (22) at (4.25, 3) {};
		\node[style=circle,fill=black] (23) at (3.5, 2) {};
		\node [style=circle,fill=black] (24) at (2.75, 1) {};
		\node[style=circle,fill=black] (25) at (2, 0) {};
		\draw[]  (1,-0.4) rectangle (3,0.4);
\node at (2,-0.8) {base solver};
		\node [style=circle,fill=black] (26) at (1.25, 1) {};
		\node [style=circle,fill=black] (27) at (0.5, 2) {};
		\node[style=circle,fill=black] (28) at (-0.25, 3) {};
\draw[style=dashed] (-2,1.5) -- (6,1.5); 
\draw[style=dashed] (-2,3.5) -- (6,3.5); 
\node (p1) at (2,1) {FP64};
\node (p2) at (2,2.5) {FP32};
\node (p3) at (2,4.5) {FP16};
\node[left of=28] (a)  {$S_{32},R_{32\rightarrow32\qquad}$};
\node[left of=32] (a)  {$S_{16},R_{16\rightarrow16\qquad}$};
\node[left of=31] (a)  {$S_{16},R_{16\rightarrow32\qquad}$};
\node[left of=26] (a)  {$S_{64},R_{64\rightarrow64\qquad}$};
\node[left of=27] (a)  {$S_{32},R_{32\rightarrow64\qquad}$};

\node[right of=24]  (a)   {$\qquad S_{64},P_{64\rightarrow32}$};
\node[right of=33] (a)  {$\qquad S_{16},P_{16\rightarrow16}$};
\node[right of=30] (a)  {$\qquad S_{16},P_{16\rightarrow16}$};
\node[right of=22] (a)  {$\qquad S_{32},P_{32\rightarrow16}$};
\node[right of=23] (a)  {$\qquad S_{32},P_{32\rightarrow32}$};
\node (h1) at (9,5) {level 5};
\node (h1) at (9,4) {level 4};
\node (h1) at (9,3) {level 3};
\node (h1) at (9,2) {level 2};
\node (h1) at (9,1) {level 1};
\node (h1) at (9,0) {level 0};
		\draw (22.center) to (23.center);
		\draw (23.center) to (24.center);
		\draw (24.center) to (25.center);
		\draw (26.center) to (25.center);
		\draw (27.center) to (26.center);
		\draw (28.center) to (27.center);
	\draw (28.center) to (31.center);
	\draw (25.center) to (30.center);
	\draw (32.center) to (31.center);
	\draw (33.center) to (30.center);
\end{tikzpicture}}
 	\caption{Floating-point precision for the main operations on grid levels within a multigrid cycle \textbf{\func{HSD\_MG}}
 		($S$: smoothing, $R$: restriction, $P$: prolongation)}
 	\label{dsh_mg}
 \end{figure}

The employed overall solver is the \func{IR} with residuum scaling in Algorithm~\ref{algo:mixed_IR2} using one of the four variants of mixed-precision multigrid given in Table~\ref{variants}.
All variants use residuum calculation and solution update in \func{FP64} precision.
The double-precision variant, \textbf{\func{D\_MG}}, is used as a baseline for the performance analysis.
For \textbf{\func{H\_MG}}, the multigrid is performed entirely in \func{FP16}.
\textbf{\func{DSH\_MG}} and \textbf{\func{HSD\_MG}} use three precisions within the multigrid cycle and are reverse of each other.
\textbf{\func{HSD\_MG}} solves the fine levels in \func{FP16}, intermediate levels in \func{FP32}
and the coarsest grid (base solver) in \func{FP64}.
A V-cycle is employed in all variants. Figure \ref{dsh_mg} shows how restriction and prolongation operators change from one precision to another for the \textbf{\func{HSD\_MG}} within a cycle.
If the grid is at transition level, restriction and prolongation are performed in the precision of the current level and the result is cast to the precision of the next level. 
Casting between half- and single-precision is performed using CUDA intrinsic functions. 
For the \textbf{\func{DSH\_MG}} case, a residuum scaling is also employed when restricting to the half precision levels.
Since a linear iteration of a purely double-precision multigrid iteration would not implement the residuum scaling, we omit this step also for the variant \textbf{\func{D\_MG}} to provide a fair baseline.

The outer solver is said to be converged when the euclidean norm of the residuum vector is less than $10^{-9}$ and \func{FP64} precision is therefore mandatory for the outer \func{IR} loop. 
For the CG base solver on the coarsest level, convergence is reached when the norm of the absolute residuum 
is less than $10^{-4}$. 
As expected,
we observed that further increasing 
the accuracy of the base solver
for \textbf{\func{H\_MG}} and \textbf{\func{HSD\_MG}}
is counterproductive as precision will be lost on \func{FP16} coarse grids.
\par

\section{Implementation Details}
FEM discretization of PDEs leads to matrices with comparable numbers of non-zero elements in each row. 
For those kinds of matrix, the ELLPACK format offers significantly better performance and bandwidth than compressed sparse row (CSR) format and coordinate (COO) format \cite{NvidisSPMV}.
In our implementation, the multigrid assembles matrices for prolongation, restriction and  discretization at different grid levels.
These matrices are stored in the precision of its grid level and transferred to the device before solving the equation. The only data movement during the iteration is copying the norm of the residuum to the host.
For vector containers, the Thrust library \cite{thrust} is used which also manages the data transfer between host and device.
The execution time of the algorithm is measured by the time elapsed between the  CUDA events.
The program is executed on an accelerated compute node of the {\it UNDISCLOSED} supercomputer employing one NVIDIA V100 GPU.
Most operations used in the multigrid method are essentially variants of SpMV and all our kernels are manually written and optimized. 
We did not use libraries such as cuSPARSE \cite{cusparse} in order to have fine-grained control about the \func{FP16} and mixed-precision details of the implementation. In particular, 
kernels are fused together to minimize data access and to allow that a precision change is performed along with rescaling. E.g., line~6 and~7 of Algorithm~\ref{algo:mixed_IR2} have been implemented in one kernel \func{\small{UpdateResiduum\_Correction}} and the precision conversion and residuum scaling in line~4 of Algorithm~\ref{algo:mixed_IR2} are also performed in a single kernel. 
Similar kernels appear within the multigrid cycle.
To control the behaviors of the floating-point arithmetic, two compiler flags are used.
To enforced the fused multiply-add operation (FMA) from the IEEE 754 standard,  \func{--fmad} is enabled.
In addition, single-precision denormal values are flushed to zeros by enabling~\func{--ftz}.

\section{Analysis}

\subsection{Convergence and Performance}


For the 2D domain, we employ five numerical experiments with a grid hierarchy of 10 levels and vary the grid size from $4097^2$ to $6145^2$ degrees of freedom on the finest level. Similarly, the mesh for the 3D example ranges between $193^3$ and $289^3$ degrees of freedom.
For the model functions, we choose the values $k=1$, $k=20$ and $k=400$ in Equation~\ref{testFunc} to vary between very smooth and highly oscillating solutions.
As the parameter $k$  increases, the function
becomes more difficult to approximate due to its high-frequency oscillating nature. As initial guess for the iteration,  random values in the range [0,1] are employed for the measurements.
In the \textbf{\func{DSH\_MG}} variant, a CG base solver is solved at level \num{0} in half-precision. Multigrid operations on the level \num{1} are performed in \func{FP16}, followed by \func{FP32} on level \num{2}.
The rest of the levels is in \func{FP64}. 
\textbf{\func{HSD\_MG}} follows the inverse approach with base solver in double precision. Here, we also use one additional level of double precision, one level of single precision for transition and the rest of the levels are in half precision.

%

\begin{table}[h]
	\centering
\caption{Average iteration count until convergence 
}
 \label{itercount}
\begin{tabular}{@{}clcccc@{}}
\toprule & & \textbf{\func{D\_MG}} & \textbf{\func{H\_MG}} & \textbf{\func{HSD\_MG}} & \textbf{\func{DSH\_MG}}  \\ \midrule
\multicolumn{1}{c|}{{2D}} & $k=1$   & 13.0    & 13.2   & 13.2      & 13.2    \\ \cmidrule(l){2-6} 
\multicolumn{1}{c|}{}                    & $k=20$  & 13.0    & 14.0    & 14.0     & 14.0      \\ \cmidrule(l){2-6} 
\multicolumn{1}{c|}{}                    & $k=400$ & 13.0    & 15.0    & 15.0      & 15.0      \\ \midrule

\multicolumn{1}{c|}{{3D}} & $k=1$   & 9.5    & 13.3  & 14.8      &9.5   \\ \cmidrule(l){2-6} 
\multicolumn{1}{c|}{}                    & $k=20$  & 14.0   &14.3   & 14.8      & 14.5      \\ \cmidrule(l){2-6} 
\multicolumn{1}{c|}{}                    & $k=400$ & 7.0   & 12.8    & 14.8     & 16.3   \\ \bottomrule
\end{tabular}
\end{table}

In Table~\ref{itercount}, the averaged iteration counts over the different grid sizes are shown for different test cases and choices of the mixed-precision multigrid cycle. For all cases, the solver converges within at most 14 or 15 iterations. In particular, although the mixed-precision variants compute the correction in lower precision, the required iterations to reach the prescribed accuracy of the outer iteration is roughly the same as for \textbf{\func{D\_MG}}.
This shows that performing multigrid operations and base solver in half precision does not increase the solver iteration count significantly. 

\begin{figure}
  \centering
  \scalebox{1.0}{\includegraphics{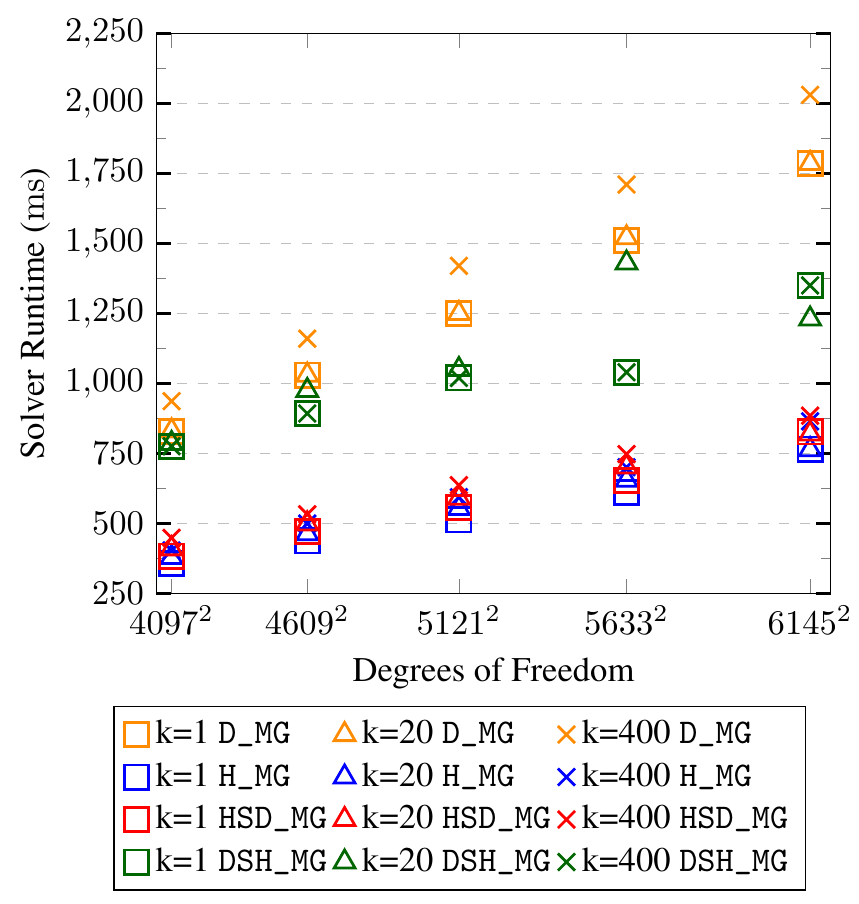}}
  \caption{Runtime of the geometric multigrid variants for the 2D domain}
  \label{timetaken}
\end{figure}

\begin{figure}
  \centering
  \scalebox{1.0}{\includegraphics{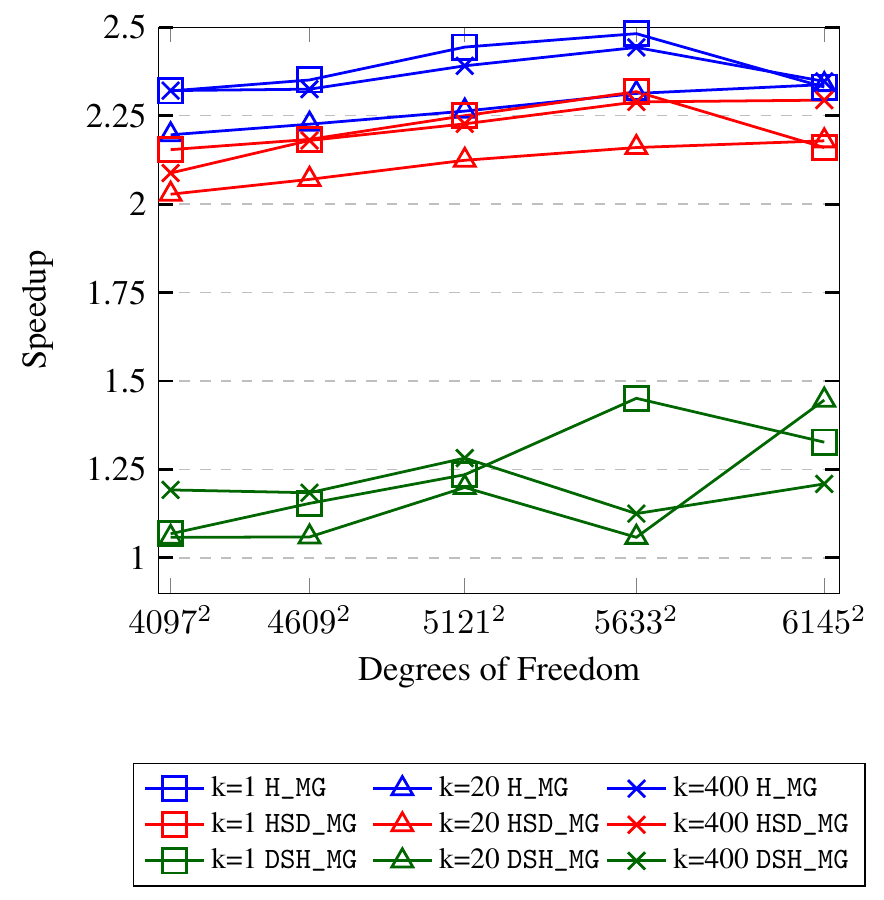}}
  \caption{Speedup of mixed-precision variants over \textbf{\func{D\_MG}} for the 2D domain}
  \label{speedup}
\end{figure}


In Figure~\ref{timetaken}, the time for the overall solver execution until convergence is presented for the 2D case. A substantial saving in runtime is observed for the half-precision variants in comparison to the double-precision baseline. Comparing both half-precision variants there is not much of a difference. This is to be expected from the fact that the iteration count is the same and the higher precision is employed for \textbf{\func{HSD\_MG}} only on very coarse meshes with few unknowns. 

In Figure~\ref{speedup}, the gained speedup of the half-precision variants over the double-precision baseline in 2D are shown. In all cases, a speedup between $2 \times$ and $2.5\times$ is achieved.
\textbf{\func{H\_MG}} performs better than  \textbf{\func{HSD\_MG}} for all $k$ cases.
A trend for better speedup with increasing problem size is observed.
Using half-precision floating-point arithmetic in higher grid levels means that the majority of the workload in multigrid preconditioning is accelerated by exploiting the performance of \func{FP16}. In the best case, a speedup of up to $2.5\times$ is achieved for \textbf{\func{H\_MG}} and a sufficiently large problem size.
\textbf{\func{DSH\_MG}} performs poorly for all $k$ values as it executes most multigrid operations in double precision, in particular on fine grids.

Most importantly, the quasi-constant iteration count is the enabling requirement to obtain the speedup. While each multigrid cycle is obviously speed up by the employment of \func{FP16} precision, an increase of the \func{IR} iteration count might be suspected and the comparable number of iterations requires the following explanation for an intuitive understanding. By recomputing the residuum in \func{FP64} after every multigrid cycle, the iteration path is kept close to a classical multigrid iteration in full precision. With every step, the error is only reduced by some orders of magnitude, commonly not more than a factor $10^{-1}$ or $10^{-2}$ for multigrid. Including a proper rescaling of the residuum and thereby also correction, this magnitude of reduction can already by captured by \func{FP16} and full precision is not required. These findings are in line with literature results \cite{Goeddeke2007221} for \func{FP32} multigrid preconditioning for an outer \func{FP64} iterative refinement where also two or three multigrid cycles at lower precision have been possible before residuum recomputation in higher precision.

\begin{figure}
  \centering
  \scalebox{1.0}{\includegraphics{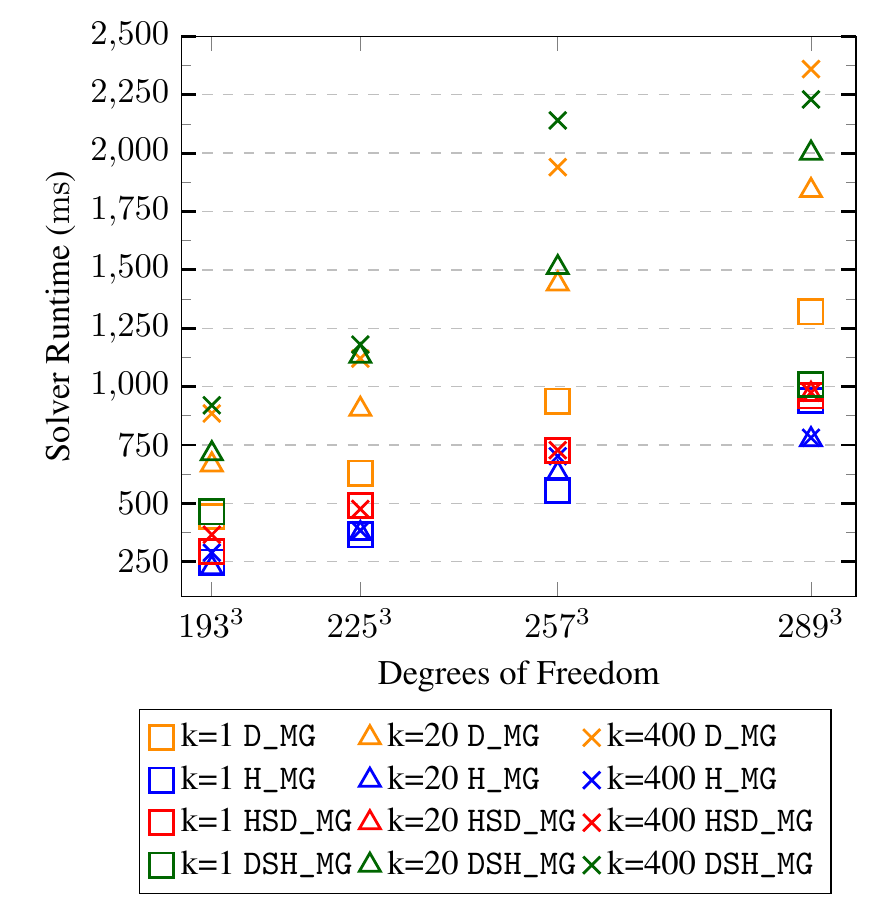}}
  \caption{Runtime of the geometric multigrid variants for the 3D domain}
  \label{timetaken3d}
\end{figure}

\begin{figure}
  \centering
  \scalebox{1.0}{\includegraphics{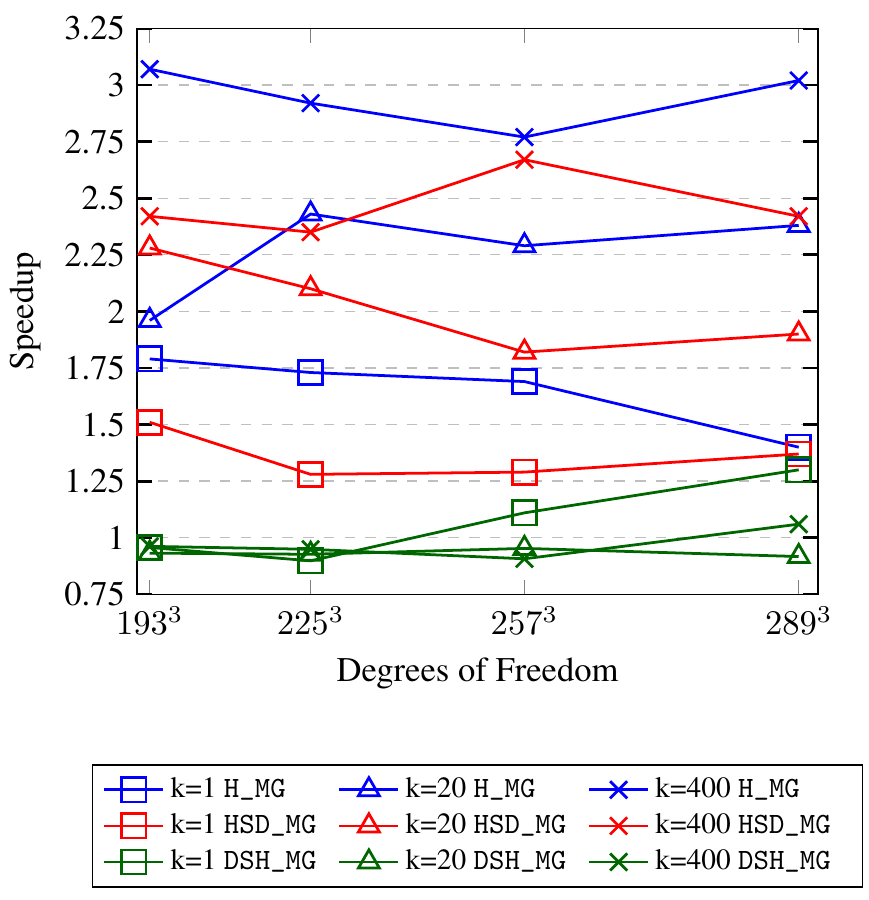}}
  \caption{Speedup of mixed-precision variants over \textbf{\func{D\_MG}} for the 3D domain}
  \label{speedup3d}
\end{figure}

In Fig.~\ref{timetaken3d}, we compare the runtimes for the 3D case. The corresponding speedup is shown in Fig.~\ref{speedup3d}. As in the 2d case, the \textbf{\func{DSH\_MG}} choice is only as good as the double precision baseline.  For the half precision variants, we can again observe a considerable speedup. However, in this example, there is some more variation in the iteration counts of the \func{IR} scheme. This leads to speedups between $1.25\times$ and $3\times$.

\subsection{Performance Limitation}

\begin{figure}
  \centering
  \scalebox{0.8}{\includegraphics{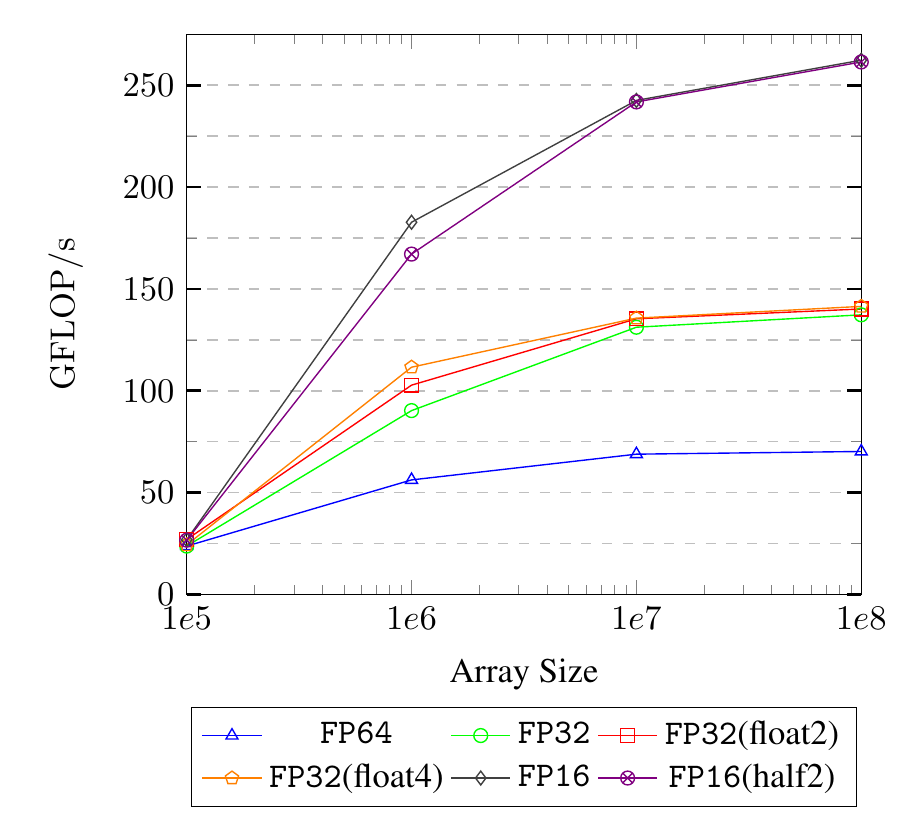}}
  \caption{Benchmarking performance of AXPY kernel in floating-point precisions supported by V100 GPU}
  \label{axpy}
\end{figure}

Our solver mainly consists of SpMV and AXPY-like kernels and we are therefore memory-bound. We thus expect a speedup proportional to the savings in memory transfer and a factor $4\times$ appears to be the upper limit. However, only a speedup of $3\times$ has been observed. We attribute this to three aspects:
\begin{itemize}
    \item We compare with the \func{FP64} baseline in which we omit the unnecessary residuum scaling. Thus, the additional residuum scaling overhead is always present for the \func{FP16} variants.

    \item The optimal speedup is only approached for very large degree of freedom numbers. As an example, we present a microbenchmarking for an AXPY kernel in Figure \ref{axpy}, where we also show the vectorized data types \func{float2}, \func{float4} and \func{half2} supported by CUDA, i.e. \func{half2} arithmetic instructions operate on two \func{FP16} values simultaneously. The vectorized loads increase throughput, reach a saturation for largest vector sizes and also show less performance for small vector sizes.
    Since a part of the execution within a multigrid cycle is performed on coarser mesh levels with fewer DoF numbers and thereby smaller vector and matrix sizes, this effect may impact the overall runtime.

    \item For some cases, the \func{FP16} variants requires more iterations. Thus, although the single iterate is faster, the speedup is reduced a bit for the time to solution since more iterations are required.
\end{itemize}

\subsection{Roofline Model for Kernels}

We employ the hierarchical Roofline model \cite{Yang_2019} to characterize our kernels. The Empirical Roofline Toolkit (ERT) \cite{ert} is used for this purpose and kernel data collected by the \func{nvprof} tool.

\begin{figure}
	\centering
\scalebox{.7}{\includegraphics[width=\textwidth]{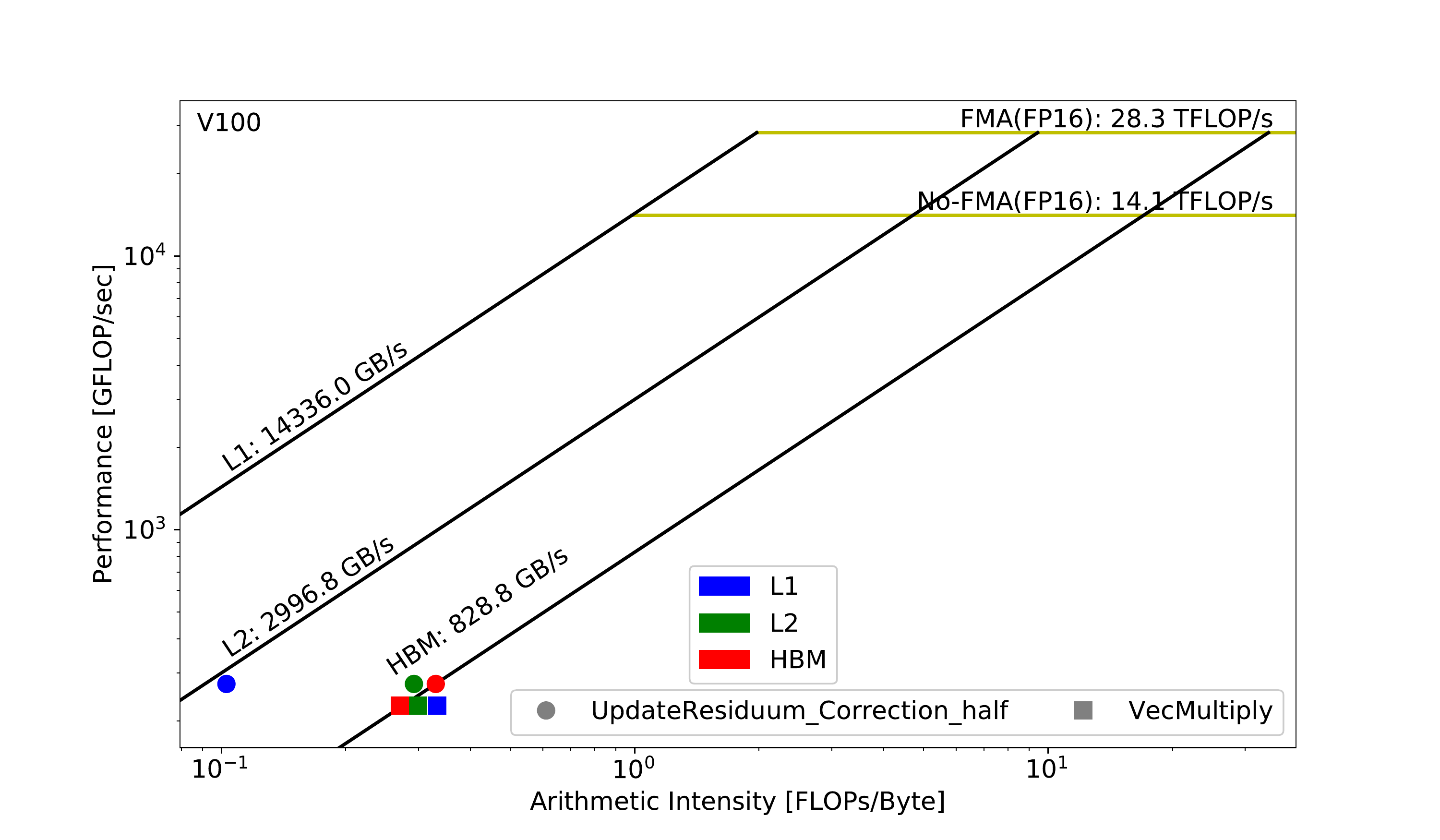}}
	\caption{Hierarchical Roofline analysis on V100 GPU}
	\label{roof}
\end{figure}

The performance of two frequently used kernels,  \func{\small{UpdateResiduum\_Correction\_half}} and \func{\small{VecMultiply}}, are analyzed. 
The former  is a fused kernel that updates both residuum  and current iterate vector.
It updates the residuum by performing a SpMV routine and adds the computed correction.
The latter kernel simply performs a component-wise vector multiplication and is used for the Jacobi smoothing.
In Figure~\ref{roof}, the measured results are depicted. 

For the \func{\small{UpdateResiduum\_Correction\_half}} kernel, L1~(blue) and L2~(green) intensities are widely separated, and L2 and HBM~(red) are very close.
This implies that substantial data reuse comes from L1 cache or in register reuse and there is virtually no data reuse in the L2 cache.
HBM intensity is larger than L1 intensity which means that there is much higher locality in the caches than in the register file.
This may be the benefit of caching the vector of SpMV routine in read-only data cache by using\func{ \_\_ldg()} intrinsic function.
However, it does not support \func{FP16} yet so implicit type casting to float occurs.
The kernel is clearly not bounded by either L1 or L2 bandwidth but by HBM.
This can also be explained by the fact that SpMV operations are memory-bound. 

The \func{VecMultiply} kernel performs one floating-point operation on each element of two input vectors and a write operation on the element of a third vector.
It is therefore also very memory-bound, which can be inferred from the Roofline model. 
In Figure \ref{roof}, L1, L2, and HBM intensities are very close which implies that there is virtually no data reuse and its performance is limited by device memory bandwidth.

\section{Conclusion}
We investigated the benefit of half-precision accuracy for geometric multigrid preconditioning within iterative refinement on the NVIDIA V100 GPU. The solution for a sparse linear system of equations stemming from the finite element method can be accelerated employing the mixed-precision approach. Using one half-precision multigrid V-cycle to approximate the correction equation of the iterative refinement, the iteration count of the iterative refinement is not much affected but a speedup of up to $3 \times$over the double-precision multigrid is observed for the overall solution time since each iterate is speed up by the employment of half precision in the memory-bound computation.


\par
\par

  \section{Acknowledgment}
 The authors gratefully acknowledge the Gauss Centre for Supercomputing e.V. (www.gauss-centre.eu) for funding this project by providing computing time on the GCS Supercomputer JUWELS at Jülich Supercomputing Centre (JSC). 


\clearpage 
\bibliographystyle{unsrt}
\bibliography{Oo-Vogel-FP16-MG}  

\end{document}